\documentclass[runningheads]{llncs}
\usepackage[T1]{fontenc}
\usepackage{graphicx}
\usepackage{marvosym}
\usepackage{booktabs}
\usepackage{multirow}
\usepackage{makecell}
\usepackage{hyperref}

\usepackage{bbding}
\begin{document}
\titlerunning{Automated Detection of RDR and DME}
\title{Deep Learning-Based Detection of Referable Diabetic Retinopathy and Macular Edema Using Ultra-Widefield Fundus Imaging}
\author{Philippe Zhang\inst{1,2,3}
\and
Pierre-Henri Conze\inst{1,4}
\and
Mathieu Lamard\inst{1,2}
\and
Gwenolé Quellec\inst{1}
\and
Mostafa El Habib Daho\inst{1,2}\textsuperscript{(\Letter)}}
\authorrunning{P. Zhang et al.}
\institute{
LaTIM UMR 1101, Inserm, Brest, France \and
Univ Bretagne Occidentale, Brest, France \and
Evolucare Technologies, Villers-Bretonneux, France \and
IMT Atlantique, Brest, France \\
\email{mostafa.elhabibdaho@univ-brest.fr}}
\maketitle
\begin{abstract}
Diabetic retinopathy and diabetic macular edema are significant complications of diabetes that can lead to vision loss. Early detection through ultra-widefield fundus imaging enhances patient outcomes but presents challenges in image quality and analysis scale. This paper introduces deep learning solutions for automated UWF image analysis within the framework of the MICCAI 2024 UWF4DR challenge. We detail methods and results across three tasks: image quality assessment, detection of referable DR, and identification of DME. Employing advanced convolutional neural network architectures such as EfficientNet and ResNet, along with preprocessing and augmentation strategies, our models demonstrate robust performance in these tasks. Results indicate that deep learning can significantly aid in the automated analysis of UWF images, potentially improving the efficiency and accuracy of DR and DME detection in clinical settings.

\end{abstract}

\section{Introduction}
Diabetic retinopathy (DR) and diabetic macular edema (DME) are among the leading causes of preventable blindness worldwide, predominantly affecting the working-age population~\cite{ref_who}. DR is characterized by damage to retinal blood vessels, leading to vision impairment and potential blindness if left untreated. DME involves fluid accumulation in the macula, further exacerbating visual deterioration.

Ultra-widefield (UWF) fundus imaging is a revolutionary diagnostic tool that captures a comprehensive view of the retina, revealing peripheral lesions that standard fundus photography might miss. This extended field of view allows for better detection of peripheral lesions and earlier identification of DR and DME \cite{ref_uwf_advantages}.

Despite the advantages, the clinical adoption of UWF imaging faces significant challenges. The primary issues include the variability in image quality due to patient cooperation, imaging conditions, and the inherent complexity of interpreting wide-field images, which often require considerable expertise. 

Automated analysis of UWF fundus images using artificial intelligence (AI) presents promising advantages by potentially reducing the time and expertise needed to interpret these images, thereby increasing the scalability of DR and DME screening programs \cite{ElHabibDaho2023,ElHabibDaho2024}.

Recent advancements in deep learning, particularly in the field of convolutional neural networks (CNNs) and transformers, have shown remarkable success in image recognition tasks \cite{He2023Transformers,Li2023,ElHabibDaho2024DISCOVER}. Networks such as EfficientNet \cite{ref_efficientnet} and ResNet \cite{ref_resnet} have set new benchmarks in accuracy and efficiency, making them ideal candidates for medical image analysis tasks. Moreover, strategies like transfer learning, ensemble learning, and test-time augmentation have further enhanced their performance, particularly in scenarios with limited annotated medical imaging data \cite{Li2024,Zhang2024}.

This paper discusses our approach to leveraging these technologies to address three tasks in the analysis of UWF images as part of the MICCAI 2024 UWF4DR challenge: assessing image quality, identifying referable DR, and detecting DME. Each task presents unique challenges and requires tailored solutions, from preprocessing techniques to model architecture choices. By integrating these methods, we aim to demonstrate the efficacy of deep learning in improving the diagnostic capabilities of UWF fundus imaging, thus supporting ophthalmologists and enhancing patient care outcomes.

\section{Task 1: Image Quality Assessment}

\subsection{Dataset}
\begin{figure}
    \centering
    \includegraphics[width=0.85\linewidth]{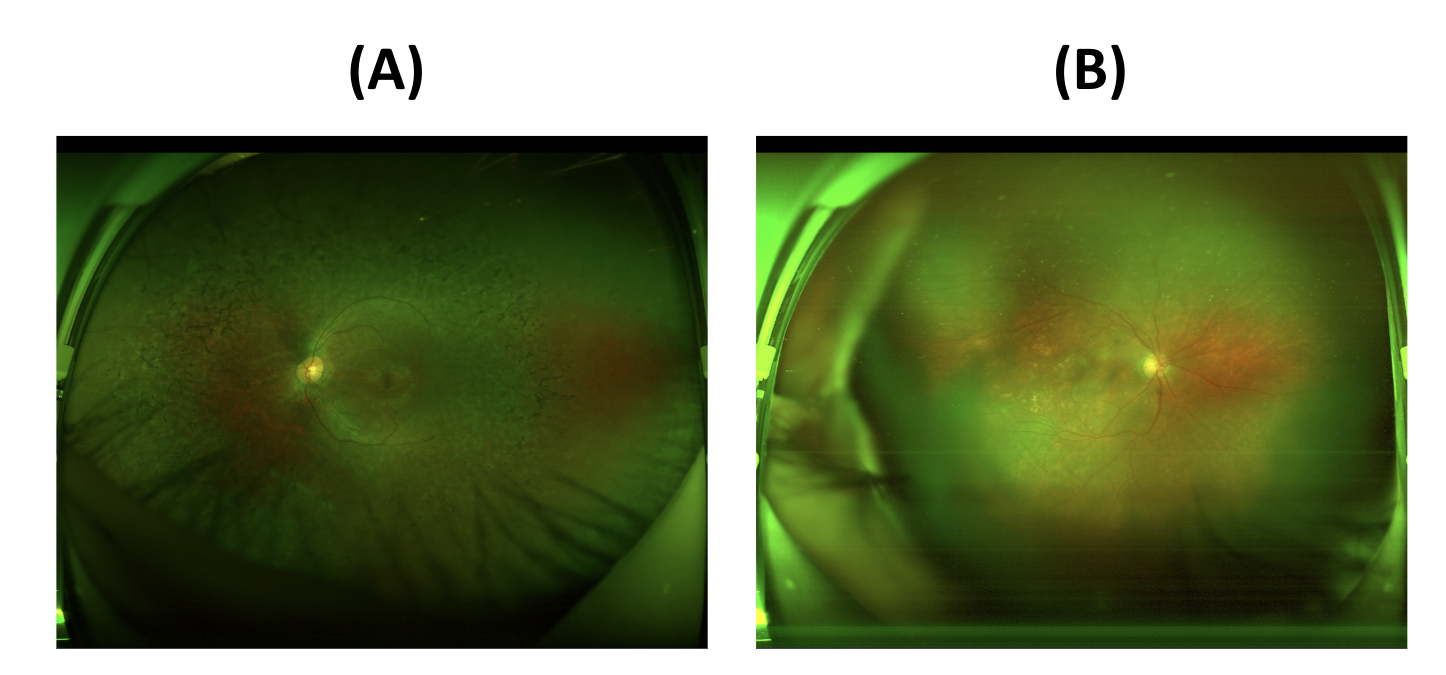}
    \caption{Samples from the quality assessment dataset: (A) good quality, (B) bad quality.}
    \label{fig:QA}
\end{figure}

This study employed the UWF fundus imaging datasets provided for the MICCAI 2024 UWF4DR challenge, specifically designed for assessing the quality of fundus photographs. The training dataset comprised 434 images, and the validation dataset included 61 images, allowing for a robust evaluation of the model's performance. Each image in these datasets was annotated with a binary label indicating the quality of the image: 1 for good quality Fig.\ref{fig:QA}(A) and 0 for bad quality Fig.\ref{fig:QA}(B). This binary classification facilitates a focused approach to training our models to distinguish between usable and non-usable images for clinical assessment and diagnosis.

\subsection{Materials and Methods}

\subsubsection{Preprocessing}

\begin{itemize}
    \item \textbf{Image Cropping and Resizing}\\
    The initial preprocessing step adjusts the original UWF fundus images, which are dimensioned at 1016$\times$800 pixels. To standardize the input for uniform analysis, these images are center-cropped to an 800$\times$800 pixel format (Fig.\ref{fig:normalized}(B)). This cropping focuses on the central retinal area, removing less informative peripheral regions and ensuring consistency across all processed images. The cropped images were then resized to 448$\times$448 pixels, a dimension determined optimal for maintaining sufficient detail while allowing efficient processing.

    \item \textbf{Color Normalization}\\
    Given the intrinsic variability in fundus imaging conditions, robust color normalization is essential for standardizing image inputs to our deep learning models. Our methodology involves a sophisticated color normalization technique that adapts to the unique characteristics of each image, enhancing the model's focus on textural and structural integrity rather than mere color variations.

    We implemented a local mean subtraction technique using the Python Imaging Library (PIL), which operates by adjusting each color channel of the RGB images independently. This process entails the following steps:
    \begin{itemize}
        \item \textbf{Gaussian Blurring:} Each color channel (Red, Green, Blue) of the image is subjected to Gaussian blurring, a smoothing technique that helps in reducing high-frequency noise components. This blurring is parameterized by a radius that dictates the extent of smoothing, effectively creating a blurred version of the original image that serves as an estimate of the local mean color.
        \item \textbf{Local Mean Subtraction:} The blurred image is then subtracted from the original image to highlight deviations from the local mean. This step enhances local contrast and emphasizes edges and fine details, which are crucial for accurate image quality assessment.
        \item \textbf{Amplification and Offset Adjustment:} After subtracting the blurred image, the result is amplified to increase the dynamic range of the processed image. An offset is then added to ensure that all pixel values remain within the valid range of [0, 255]. This step ensures that the resultant image maintains a balanced brightness and contrast level, which is suitable for further analysis.
        \item \textbf{Channel Reintegration:} The individually processed channels are recombined to form the final, color-normalized image (Fig.\ref{fig:normalized}(C)).
    \end{itemize}

    \begin{figure}
        \centering
        \includegraphics[width=1\linewidth]{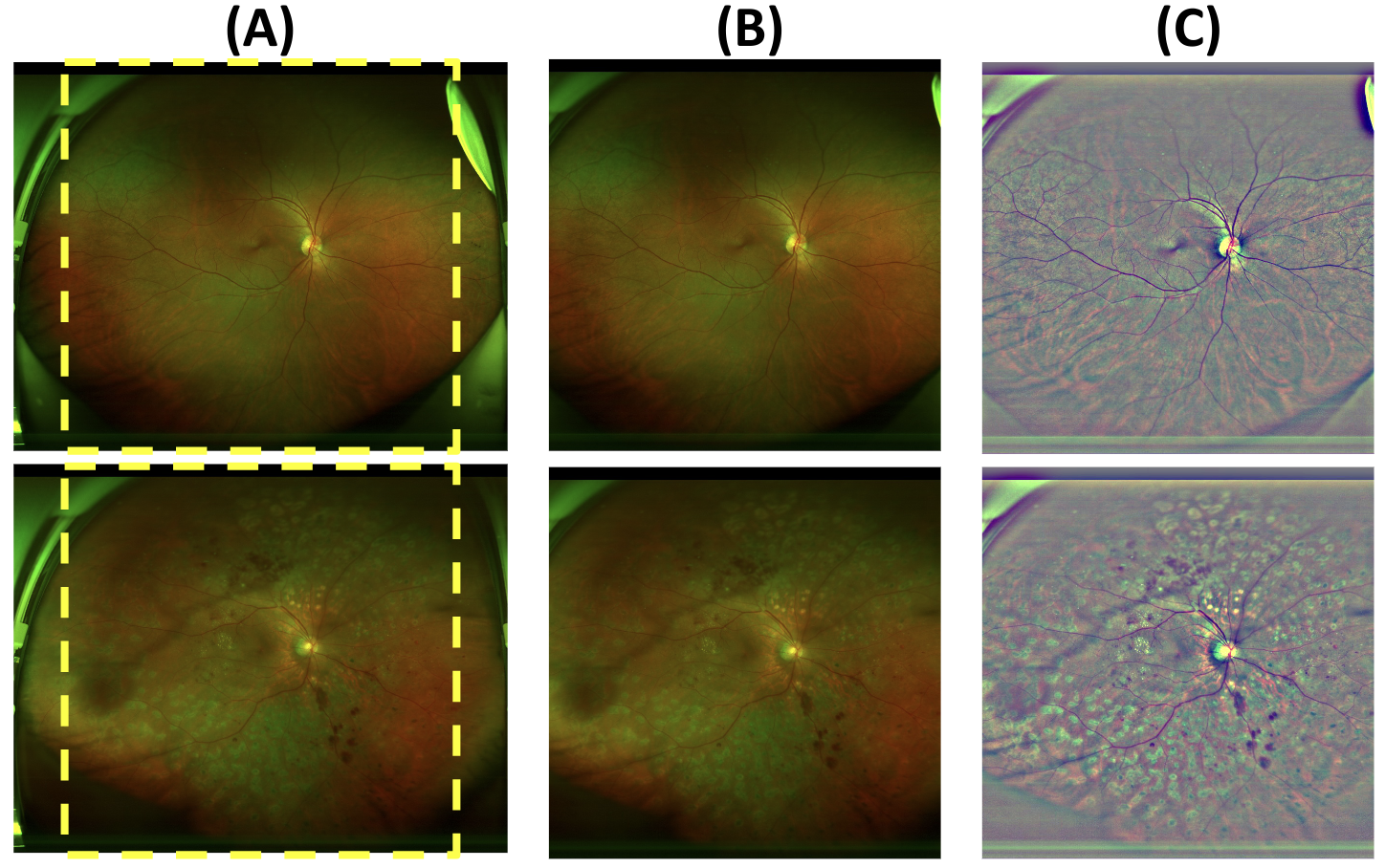}
        \caption{Preprocessing of images: (A) original image, (B) center-cropped image, (C) preprocessed image.}
        \label{fig:normalized}
    \end{figure}

    \item \textbf{Data Augmentation}\\
    To improve model robustness, we employed data augmentation strategies during training, including random horizontal and vertical flips, rotations (up to 45 degrees), brightness and contrast adjustments, and random zooms. These augmentations help the model generalize better by exposing it to various image transformations.

\end{itemize}

\subsubsection{Model Architecture}
For robust image quality assessment, we engineered an ensemble of three EfficientNet-based models, each tailored to capture different aspects of image quality across varying scales and complexities:

\begin{enumerate}
    \item \textbf{EfficientNet-B0 (800$\times$800 crop)}: This model is fine-tuned on images cropped to 800$\times$800 pixels and resized to 448$\times$448, a dimension chosen to maintain a balance between detail retention and computational efficiency. The model utilizes the standard EfficientNet-B0 architecture~\cite{ref_efficientnet}, known for its scalable and efficient convolutional network backbone.

    \item \textbf{EfficientNet-B0 (500$\times$500 crop)}: Optimized for processing images cropped to 500$\times$500 pixels (instead of 800$\times$800) and resized to 448$\times$448, this model focuses on the core, most informative parts of the images, emphasizing essential features over peripheral details. The smaller crop size aids in faster processing times and reduces the model’s susceptibility to noise and distortions prevalent in the outer regions of the images.

    \item \textbf{Multilevel EfficientNet-B0 (800$\times$800 crop)}: The Multilevel EfficientNet-B0 (ML-EfficientNet-B0) is a custom adaptation of the standard EfficientNet-B0 model, uniquely designed to improve sensitivity to subtle nuances in image quality by integrating feature maps from different stages of the network. This model extracts feature maps from early, intermediate, and deep layers, capturing fine-grained details and higher-level abstractions crucial for comprehensive assessment (Fig.\ref{fig:ml_b0}).

    \begin{figure}
        \centering
        \includegraphics[width=1\linewidth]{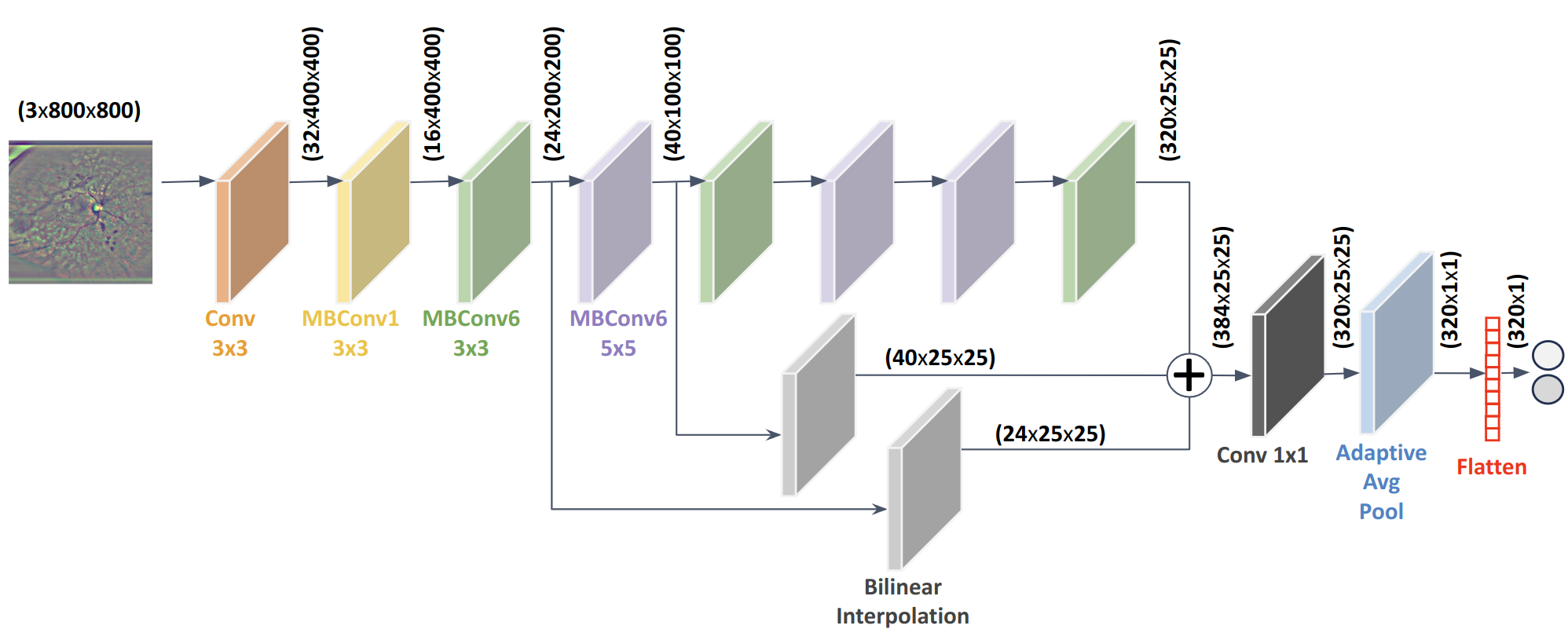}
        \caption{Architecture of the Multilevel EfficientNet-B0 model.}
        \label{fig:ml_b0}
    \end{figure}

\end{enumerate}

\subsubsection{Training}
All models were initialized with ImageNet pre-trained weights and trained for 100 epochs using the Adam optimizer with an initial learning rate of $1e^{-4}$ and an ExponentialLR scheduler. Early stopping based on validation loss was employed to prevent overfitting. Cross-entropy loss was used as the loss function. The training was executed on a computer with an A6000 Ada NVIDIA GPU (48GB of VRAM). 

\subsubsection{Ensemble Prediction}
During inference, predictions from the three models were averaged to produce the final probability score, leveraging the strengths of each model.

\subsection{Results and Discussion}
Our ensemble model achieved the highest performance among the methods tested, demonstrating superior capability in assessing image quality (Table~\ref{tab:task1_results}).\\
To evaluate the models, we used Sensitivity, Specificity, AUROC, and AUPRC. Sensitivity (also known as recall) measures the proportion of true positives correctly identified by the model, while Specificity quantifies the proportion of true negatives correctly classified. The AUROC (Area Under the Receiver Operating Characteristic Curve) measures the ability of the model to distinguish between classes, with a value of 1 indicating perfect classification and 0.5 representing random chance. The AUPRC (Area Under the Precision-Recall Curve) highlights the trade-off between precision (the proportion of true positives among predicted positives) and recall, particularly valuable for imbalanced datasets.\\
On the validation set, the ensemble achieved an AUROC of 0.8739 and an AUPRC of 0.8946. When evaluated on the test set, the ensemble's performance improved significantly, achieving an AUROC of 0.9051 and an AUPRC of 0.9410, indicating better generalization to unseen data.

\begin{table}[h]
\caption{Performance Metrics for Task 1}\label{tab:task1_results}
\centering
\begin{tabular}{lccccc}
\hline
\textbf{Method} & \textbf{Dataset} & \textbf{AUROC} & \textbf{AUPRC} & \textbf{Sensitivity} & \textbf{Specificity} \\
\hline
EfficientNet-B0 (crop 500) & Validation & 0.8305 & 0.8546 & \textbf{0.9459} & 0.7083 \\
EfficientNet-B0 (crop 800) & Validation & 0.8559 & 0.8910 & 0.8649 & 0.7500 \\
ML-EfficientNet-B0 & Validation & 0.8626 & 0.9073 & 0.7297 & \textbf{0.9167} \\
Ensemble Model & Validation & \textbf{0.8739} & \textbf{0.8946} & 0.7568 & 0.8750 \\
\hline
EfficientNet-B0 (crop 800) & Test & 0.8758  & 0.9205  & 0.6271  & \textbf{0.9500 } \\
ML-EfficientNet-B0 & Test & 0.8805  & 0.9217  & \textbf{0.7119}  & 0.9000 \\
Ensemble Model & Test & \textbf{0.9051} & \textbf{0.9410} & \textbf{0.7119} & \textbf{0.9500} \\
\hline
\end{tabular}
\end{table}

\subsubsection{Discussion}
The ensemble approach outperformed individual models on the validation dataset, indicating that combining diverse models enhances the ability to assess image quality accurately. The Multilevel EfficientNet-B0 model also showed strong performance, especially in specificity. However, its lower sensitivity indicates that it may miss some high-quality images. 

The ensemble model balances both sensitivity and specificity, making it more reliable for clinical use. Notably, the ensemble model demonstrated improved AUROC and AUPRC scores on the test set compared to the validation set, suggesting better generalization to unseen data. This improvement underscores the robustness of the ensemble model in real-world applications, where it is crucial to maintain high performance on diverse and previously unseen images.

The high specificity (95\%) achieved by the ensemble on the test set is particularly significant for clinical applications. A high specificity ensures that the model is effective at correctly identifying low-quality images, reducing the risk of using suboptimal images for diagnosis, which could lead to misdiagnosis or overlooked conditions.

The improved performance on the test set demonstrates that our preprocessing techniques, model architectures, and training strategies contribute to models that generalize well beyond the training data. This generalization is critical for deployment in varied clinical settings, where imaging conditions and patient populations may differ from the training data.

\section{Task 2: Identification of Referable Diabetic Retinopathy}

\subsection{Dataset}
For the identification of referable diabetic retinopathy (RDR), we utilized a combined dataset comprising images from the UWF4DR challenge and the publicly available DeepDRiD dataset~\cite{ref_deepdrid}. The UWF4DR dataset provided 201 training images and 50 validation images, each labeled for the presence or absence of RDR. The DeepDRiD dataset added diversity and volume to our training data, containing high-resolution retinal images with detailed annotations for DR severity levels ranging from no DR to proliferative DR. To binarize this dataset, we have attributed the RDR label to images that were annotated Severe Non-Proliferative Diabetic Retinopathy (NPDR) or Proliferative Diabetic Retinopathy (PDR) in the DeepDRiD dataset according to The American Academy of Ophthalmology (AAO) guidelines on diabetic retinopathy management \cite{FLAXEL2020P66}. 

By integrating these two datasets, we aimed to enhance the model's generalizability and robustness. The combined dataset exposed the model to a wide variety of imaging conditions, retinal pathologies, and demographic variations, which is critical for developing a model capable of performing well across different clinical settings.

\begin{figure}[h!]
    \centering
    \includegraphics[width=0.85\linewidth]{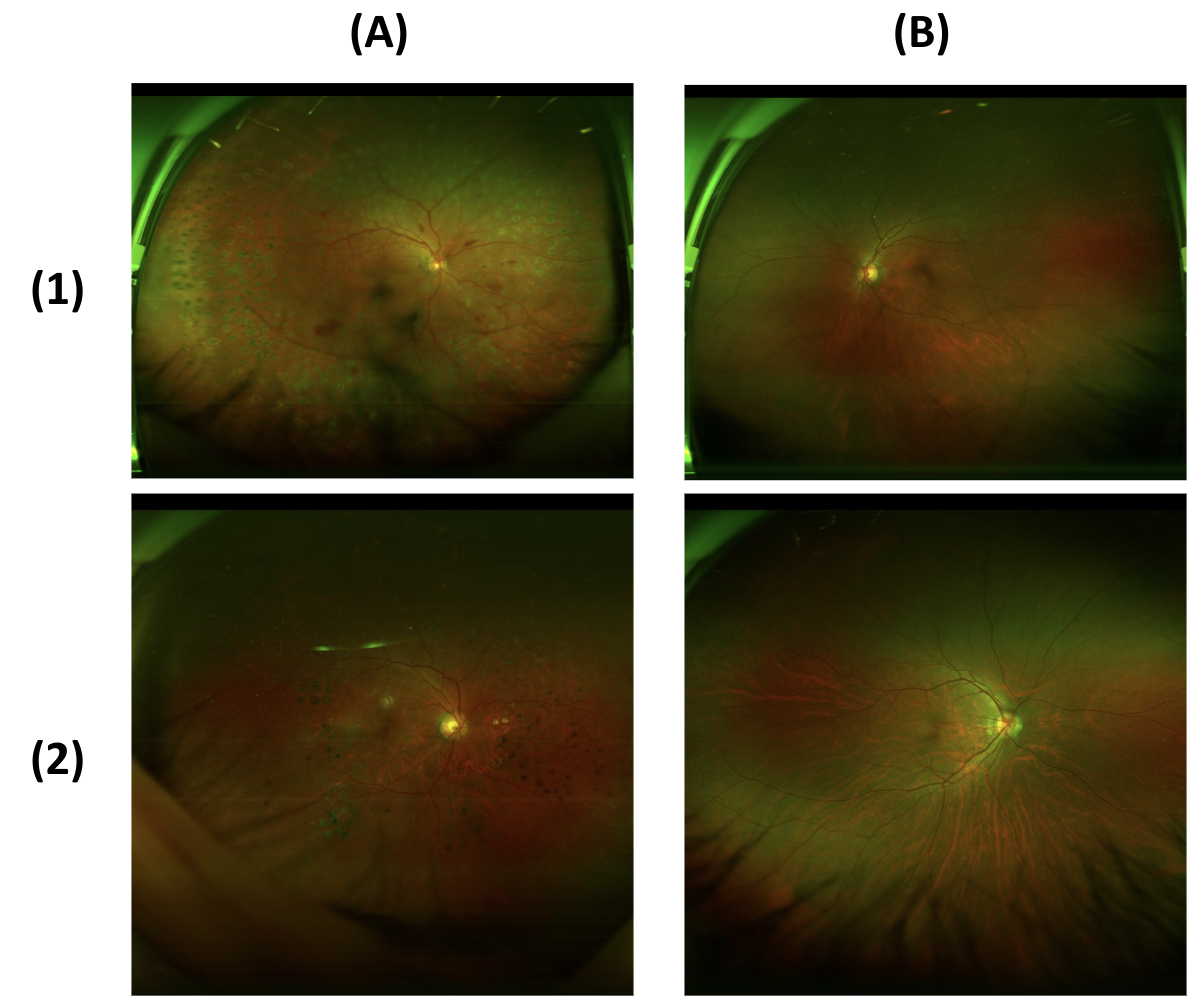}
    \caption{Examples from the RDR dataset: (A) Images with referable DR, (B) Images without referable DR, (1) Images from the UWF4DR dataset, (2) Image from the DeepDRiD dataset.}
    \label{fig:RDR_samples}
\end{figure}

Figure~\ref{fig:RDR_samples} illustrates examples of images classified as RDR and non-RDR from the dataset.

To rigorously evaluate our models and enhance their generalizability, we implemented a 5-fold cross-validation strategy on the combined dataset. This approach involved partitioning the dataset into five equal subsets, training the models on four subsets, and validating on the remaining subset in each iteration. This method ensures that every image is used for both training and validation, providing a robust assessment of model performance across different data splits.

\subsection{Materials and Methods}

\subsubsection{Preprocessing}

To ensure consistency across tasks and to leverage the preprocessing benefits observed in Task 1, we applied the same preprocessing pipeline:

\begin{itemize}
    \item \textbf{Image Cropping}: Images were center-cropped to 800$\times$800 pixels, focusing on the central retinal region where signs of DR are most prevalent. Unlike Task 1, we did not resize the images after cropping, preserving the original resolution to maintain fine details critical for detecting microaneurysms, hemorrhages, and other DR lesions.
    \item \textbf{Color Normalization}: The same local mean subtraction technique using Gaussian blurring was applied (see Fig.~\ref{fig:normalized}). This normalization enhanced contrast and highlighted pathological features, aiding the models in learning relevant patterns.
    \item \textbf{Data Augmentation}: The same as Task 1.
\end{itemize}

\subsubsection{Model Architecture}

To address the complexity of RDR identification, we proposed the following models:

\begin{enumerate}

    \item \textbf{ResNet-18}: This model is the ResNet-18 architecture~\cite{ref_resnet}, a deep network known for its strong feature extraction capabilities due to its 18 layers and residual connections, fine-tuned on the combined datasets (UWF4DR and DeepDRiD).
    
    \item \textbf{EfficientNet-B0}: This model is the EfficientNet-B0 architecture~\cite{ref_efficientnet} fine-tuned on the combined datasets (UWF4DR and DeepDRiD).

    \item \textbf{ML-EfficientNet-B0}: The same custom EfficientNet-B0 model~\cite{ref_efficientnet} proposed in Task 1 that concatenates feature maps from multiple intermediate layers (Fig.~\ref{fig:ml_b0}). This multilevel feature extraction allows the model to capture both low-level details (e.g., microaneurysms) and high-level contextual information (e.g., neovascularization patterns). The concatenated features are then passed through fully connected layers for binary classification.

    \item \textbf{ResNet-18 Ensemble}: We utilized the ResNet-18 architecture~\cite{ref_resnet}. The depth of ResNet-18 allows it to capture more complex patterns and features, which is beneficial for detecting subtle signs of RDR.\\
    Our approach involved performing 5-fold cross-validation on the combined dataset. After the cross-validation process, we evaluated the performance of the five models and selected the three best-performing models based on their validation AUROC scores. These three models were then used to form an ensemble. By ensembling the top models, we aimed to reduce variance, mitigate overfitting, and improve generalization to unseen data.

\end{enumerate}

\subsubsection{Training }

All models were initialized with ImageNet pre-trained weights to leverage learned features from a large dataset. The training was conducted for 100 epochs using the Adam optimizer with an initial learning rate of $1e^{-4}$ and an ExponentialLR scheduler. Early stopping based on validation loss was employed to prevent overfitting.

\subsection{Results and Discussion}

\subsubsection{Results}

Our models were evaluated on both the validation and test datasets. The performance metrics are summarized in Table~\ref{tab:task2_results}. 

\begin{table}[h!]
\caption{Performance Metrics for Task 2}\label{tab:task2_results}
\centering
\begin{tabular}{lccccc}
\hline
\textbf{Method} & \textbf{Dataset} & \textbf{AUROC} & \textbf{AUPRC} & \textbf{Sensitivity} & \textbf{Specificity} \\
\hline
ResNet-18 & Validation & 0.9655 & 0.9764 & 0.7931 & \textbf{1.0000} \\
EfficientNet-B0 & Validation & 0.9540 & 0.9764 & 0.8966 & \textbf{1.0000} \\
ML-EfficientNet-B0 & Validation & 0.9754 & 0.9814 & \textbf{1.0000} & 0.8571 \\
ResNet-18 Ensemble & Validation & \textbf{0.9786} & \textbf{0.9874} & 0.9655 & 0.9523 \\
\hline
ResNet-18 & Test & 0.9667 & 0.9678 & 0.8448 & \textbf{1.0000} \\
EfficientNet-B0 & Test & 0.9733 & 0.9778 & 0.9138 & 0.9578 \\
ML-EfficientNet-B0 & Test & \textbf{0.9811} & 0.9761 & 0.9483 & 0.9437\\
ResNet-18 Ensemble & Test & 0.9796 & \textbf{0.9838} & \textbf{0.9655} & 0.9437 \\

\hline
\end{tabular}
\end{table}

\subsubsection{Discussion}

The ML-EfficientNet-B0 model demonstrated the highest AUROC on the test set (0.9811), indicating that it was best at distinguishing between RDR and non-RDR cases. This makes it a strong choice for maximizing overall classification performance. However, the ResNet-18 Ensemble had the highest AUPRC (0.9838) and sensitivity (0.9655), making it particularly useful in clinical scenarios where missing positive cases (false negatives) is a critical concern. The ability of the ResNet-18 Ensemble to capture true positives with fewer missed cases emphasizes its robustness in detecting RDR. 

EfficientNet-B0 also performed well, achieving a solid AUROC (0.9733) and maintaining a good balance between sensitivity (0.9138) and specificity (0.9578). This model may be preferred in contexts where a balance between identifying true positives and avoiding false positives is important.

The ResNet-18 model demonstrated perfect specificity (1.0000) on the test set, ensuring that it does not misclassify any non-RDR cases. This is a valuable trait for minimizing false positives and reducing unnecessary follow-up interventions, although its lower sensitivity (0.8448) compared to other models suggests that it may miss some cases of RDR.

The strong performance of our models across both the validation and test sets indicates good generalization to unseen data. The high specificity and sensitivity across the models suggest that our models are capable of effectively identifying both true positives and true negatives, ensuring reliable screening outcomes.

\section{Task 3: Identification of Diabetic Macular Edema}

\subsection{Dataset} We used the UWF4DR dataset for DME detection, utilizing labels provided by the challenge organizers. The dataset includes UWF images labeled for the presence or absence of DME.

\subsection{Materials and Methods}

\subsubsection{Preprocessing} The same preprocessing pipeline from previous tasks was applied to ensure consistency.

\subsubsection{Model Architecture} In this task, we fine-tuned the models from Task 2 for DME detection, leveraging the high correlation between DR and DME~\cite{ref_dr_dme_correlation}. Specifically, we fine-tuned the EfficientNet-B0 and Multi-Level EfficientNet-B0 models for DME detection, while also utilizing Test-Time Augmentation (TTA) during inference for one model to improve performance.

\subsubsection{Training and TTA} The models were initialized with weights from Task 2. The training was conducted using the Adam optimizer with an initial learning rate of $1e^{-4}$ and an ExponentialLR scheduler. For the model utilizing TTA, multiple augmented versions of each image were generated during inference, and the predictions were averaged to obtain the final result.

\subsection{Results and Discussion}

Our models were evaluated on both validation and test datasets, with the performance metrics summarized in Table~\ref{tab:task3_results}. The Multi-Level EfficientNet-B0 with TTA achieved the highest AUROC (0.9820) and AUPRC (0.9699) on the test set, outperforming both the standard EfficientNet-B0 and the Multi-Level EfficientNet-B0 models without TTA. Notably, the TTA model also achieved the highest specificity (0.9577), crucial for minimizing false positives in clinical settings. The EfficientNet-B0 model achieved the highest sensitivity (0.9500), suggesting it is effective at correctly identifying positive cases of DME, though at a slight trade-off in specificity.

\begin{table}[h!]
\caption{Performance Metrics for Task 3}\label{tab:task3_results}
\centering
\begin{tabular}{lccccc}
\hline
\textbf{Method} & \textbf{Dataset} & \textbf{AUROC} & \textbf{AUPRC} & \textbf{Sensitivity} & \textbf{Specificity} \\
\hline
EfficientNet-B0 & Validation & 0.9425 & 0.9157 & \textbf{0.9167} & 0.9523 \\
ML-EfficientNet-B0 & Validation & 0.9445 & 0.9674 & 0.8750 & \textbf{1.0000} \\
ML-EfficientNet-B0 TTA & Validation & \textbf{0.9663} & \textbf{0.9786} & 0.8750 & \textbf{1.0000} \\
\hline
EfficientNet-B0 & Test & 0.9704 & 0.9698 & \textbf{0.9500} & 0.9436 \\
ML-EfficientNet-B0 & Test & 0.9775 & 0.9616 & 0.9250 & 0.9437\\
ML-EfficientNet-B0 TTA & Test & \textbf{0.9820} & \textbf{0.9699} & 0.9250 & \textbf{0.9577} \\

\hline
\end{tabular}
\end{table}

\subsubsection{Discussion}

The fine-tuning from Task 2 models effectively leveraged the shared features between DR and DME, leading to improved detection performance in DME detection. The Multi-Level EfficientNet-B0 model with TTA excelled in terms of AUROC and AUPRC, demonstrating that multi-level feature extraction combined with test-time augmentation enhances the model's ability to detect subtle signs of DME. This approach improved sensitivity and specificity while also reducing computational time, suggesting that it is efficient and suitable for clinical deployment.

The EfficientNet-B0 model, with the highest sensitivity (0.9500), ensures the identification of most positive DME cases, though it sacrificed some specificity compared to the Multi-Level EfficientNet-B0 with TTA. In contrast, the Multi-Level EfficientNet-B0 with TTA achieved a balance between sensitivity (0.9250) and specificity (0.9577), making it an ideal choice for minimizing false positives while maintaining a high detection rate.

Accurate detection of DME is crucial for preventing vision loss through timely intervention. The models' high sensitivity and specificity suggest they could play a vital role in DME screening programs, assisting ophthalmologists in identifying patients who require further examination and treatment.

\section{Conclusion}

In this paper, conducted within the framework of the MICCAI 2024 DWT4DR challenge, we explored the use of deep learning models for the automated analysis of UWF fundus images, focusing on three key tasks: image quality assessment, identification of referable DR, and detection of DME. By applying popular architectures such as EfficientNet and ResNet, we demonstrated robust performance across all tasks, yielding promising results that could aid in the early detection and management of diabetic eye diseases.

Our models achieved strong generalization to unseen data and effectively balanced sensitivity and specificity. The integration of multi-level feature extraction and techniques like model ensembling and test-time augmentation enhanced model robustness, making our approaches suitable for clinical deployment and potentially improving patient outcomes through early detection and management.

However, there are limitations to consider. Future work could explore multi-task learning models to simultaneously assess image quality and detect multiple pathologies (e.g., DR and DME), thereby streamlining the screening process. Additionally, a promising direction for future research is the exploration of foundation models like RetFound~\cite{RetFound}. RetFound, a self-supervised model pre-trained on a large corpus of retinal images, has the potential to enhance performance and generalization, especially in scenarios with limited labeled data. Leveraging such foundation models could facilitate multi-task learning and improve the detection of multiple pathologies within a unified framework. Further, incorporating explainability tools can improve transparency, while advanced data augmentation methods like Mixup could increase model robustness and reduce overfitting.

\end{document}